\documentclass[runningheads,a4paper]{llncs}

\usepackage{amssymb}
\usepackage{graphicx}

\usepackage{url}
\urldef{\mailsa}\path|{afrati}@softlab.ntua.gr,|
\urldef{\mailsb}\path|{vkyri}@softlab.ece.ntua.gr,|
\urldef{\mailsc}\path|{dsouliou}@mail.ntua.gr,|
\urldef{\mailsd}\path|{plekeas}@tem.uoc.gr|
\newcommand{\keywords}[1]{\par\addvspace\baselineskip
\noindent\keywordname\enspace\ignorespaces#1}

\usepackage{qtree}
\usepackage{color}

\usepackage{calc}
\usepackage{amssymb}
\usepackage{amstext}
\usepackage{amsmath}
\newcommand{\tab}{\hspace*{2em}}
\def\qed{\hspace*{\fill} $\Box$}

\newenvironment{myProof}{\noindent{\bf Proof:} }{\qed}

\begin{document}

\mainmatter  

\title{A New Framework for Join Product Skew}


%
%

\authorrunning{A New Framework for Join Product Skew}
\author{Foto Afrati\inst{1}, Victor Kyritsis\inst{1},
Paraskevas Lekeas\inst{2}, Dora Souliou\inst{1}}
\institute{National Technical University of Athens, Athens, Greece,\\
\email{\{afrati, vkyri\}@cs.ntua.gr, dsouliou@mail.ntua.gr}
\and
Department of Applied Mathematics, University of Crete, Herakleio, Greece,\\
\email{plekeas@tem.uoc.gr}
\\
}

%
%

\toctitle{Lecture Notes in Computer Science}
\tocauthor{Authors' Instructions}
\maketitle

\begin{abstract}
Different types of data skew can result in load imbalance in the context of parallel joins under the shared nothing
architecture. We study one important type of skew, join product skew (JPS). A static approach based on frequency classes is proposed which takes for granted the data distribution of join attribute values. It comes from the observation that the join selectivity can be expressed as a sum of products of frequencies of the join attribute values. As a consequence, an appropriate assignment of join sub-tasks, that takes into consideration the magnitude of the frequency products can alleviate the join product skew. Motivated by the aforementioned remark, we propose an algorithm, called Handling Join Product Skew (HJPS), to handle join product skew.

\end{abstract}

\section{Introduction}
\noindent The limited potentials of centralized database systems in terms of the storage and the process of large volumes of data has led to the advent of parallel database management systems (PDBMS) that adopt the shared-nothing architecture. According to this architecture, each computational node (database processor) has its own memory and CPU and independently accesses its local disks while it is provided with the ability to perform locally relational operations. By definition, the aforementioned architecture favors the deployment of data intensive scale computing applications \cite{mapreduce} by reducing the complexity of the underlying infrastructure and the overall cost as well.

Within the scope of the parallel evaluation of the relational operators by splitting them into many independent operators (\textit{partitioned parallelism}), sort-merge join and hash-join constitute the main algorithms for the computation of the equijoin. Equijoin is a common special case of the join operation $R \Join S$, where the join condition consists solely of equalities of the form $R.X=S.Y$ ($X$ and $Y$ are assumed to be attributes of the relations $R$ and $S$ respectively). Both algorithms are subject to parallel execution. However, the hash-based algorithm has prevailed since it has linear execution cost, and it performs better in the presence of data skew as well \cite{ParallelDatabaseSystemsThefutureofhighperformanceDatabaseSystems}.

The parallel hash-based join processing is separated into three phases. In the first phase, each relation is fully declustered horizontally across the database processors by applying a partition function on the declustering attribute, which in general is different from the join attribute. Next, at the redistribution phase, each database processor applies a common hash function $h$ on the join attribute value for its local fragments of relations $R$ and $S$. The hash function $h$ ships any tuple belonging to either relation $R$ or $S$ with join attribute value $b_i$ to the $h(b_i)$-th database processor. At the end of the redistribution process both relations are fully partitioned into disjoint fragments. Lastly, each database processor $p$ performs locally with the most cost-effective way an equijoin operation between its fragments of relations $R$ and $S$, denoted by $R^p$ and $S^p$ respectively. The joined tuples may be kept locally in each database processor instead of being merged with other output tuples into a single stream.

Skewness, perceived as the variance in the response times of the database processors involved in the previously described computation, is identified as one of the major factors that affects the effectiveness of the hash-based parallel join \cite{EffectivenessParallelJoins}. \cite{TaxonomyAndPerformanceModelDataSkewEffectsInParallelJoins} defines four types of the data skew effect: Tuple placement skew, selectivity skew, redistribution skew and join product skew. Query load balancing in terms of the join operation is very sensitive to the existence of the redistribution skew and/or the join product skew. Redistribution skew can be observed after the end of the redistribution phase. It happens when at least one database processor has received large number of tuples belonging to a specific relation, say $R$, in comparison to the other processors after the completion of the redistribution phase. This imbalance in the number of redistributed tuples is due to the existence of naturally skewed values in the join attribute. Redistribution skew can be experienced in a subset of database processors. It may also concern both the relations $R$ and $S$ (double redistribution skew). Join product skew occurs when there is an imbalance in the number of join tuples produced by each database processor. \cite{DataPlacementSharedNothingParallelDatabaseSystems} points the impact of this type of skewness to the response time of the join query. Especially, join product skew deteriorates the performance of subsequent join operation since this type of data skew is propagated into the query tree.

In this paper we address the issue of join product skew. Various techniques and algorithms have been proposed in the literature to handle this type of skew (\cite{SkewInsensitiveParallelAlgorithmsForRelationalJoin}, \cite{PracticalSkewHandlingParallelJoins}, \cite{HandlingDataSkewParallelJoinsInSharedNothingSystems}, \cite{FrequencyAdaptiveJoinForSharedNothingMachines}, \cite{DynamicJoinProductSkewHandlingHashJoinsSharedNothingDatabaseSystems}, \cite{HandlingDataSkewInParallelHashJoin}). We introduce the notion of frequency classes, whose definition is based on the product of frequencies of the join attribute values. Under this perspective we examine the cases of homogeneous and heterogeneous input relations.

We also propose a new static algorithm, called HJPS (Handling Join Product Skew) to improve the performance of the parallel joins in the presence of this specific type of skewness. The algorithm is based on the intuition that join product skew comes into play when the produced tuples associated with a specific value overbalance the workload of a processor.
HJPS algorithm constitutes a refinement of the PRPD algorithm \cite{HandlingDataSkewParallelJoinsInSharedNothingSystems} in the sense that the exact number of the needed processors is defined for each skewed value instead of duplicating or redistributing the tuples across all the database processors. Additionally, HJPS is advantageous in the case of having join product skew without having redistribution skew.

The rest of this paper is organized as follows. Section \ref{RelatedWorkSection} discusses the related work. In section \ref{ExamplesSection} we illustrate the notion of division of join attribute values into classes of frequencies by means of two generic cases. In section \ref{AlgorithmSection} an algorithm that helps in reducing join product skew effect is proposed and section \ref{ConclusionSection} concludes the paper.

\section{Related Work}\label{RelatedWorkSection}
\noindent The achievement of load balancing in the presence of redistribution and join product skew is related to the development of static and dynamic algorithms. In static algorithms it is assumed that adequate information on skewed data is known before the application of the algorithm. \cite{SkewInsensitiveParallelAlgorithmsForRelationalJoin}, \cite{PracticalSkewHandlingParallelJoins} and \cite{HandlingDataSkewParallelJoinsInSharedNothingSystems} expose static algorithms. On the contrary, \cite{FrequencyAdaptiveJoinForSharedNothingMachines}, \cite{DynamicJoinProductSkewHandlingHashJoinsSharedNothingDatabaseSystems} and \cite{HandlingDataSkewInParallelHashJoin} propose techniques and algorithms according to which data skew is detected and encountered dynamically at run time.

\cite{FrequencyAdaptiveJoinForSharedNothingMachines}, \cite{HandlingDataSkewInParallelHashJoin} address the issue of the join product skew following a dynamic approach. A dynamic parallel join algorithm that employs a two-phase scheduling procedure is proposed in \cite{HandlingDataSkewInParallelHashJoin}. The authors of \cite{FrequencyAdaptiveJoinForSharedNothingMachines} present an hybrid frequency-adaptive algorithm which dynamically combines histogram-based balancing with standard hashing methods. The main idea is that the processing of each sub-relation, stored in a processor, depends on the join attribute value frequencies which are determined by its volume and the hashing distribution.

\cite{SkewInsensitiveParallelAlgorithmsForRelationalJoin}, \cite{PracticalSkewHandlingParallelJoins} and \cite{HandlingDataSkewParallelJoinsInSharedNothingSystems} deal with the join product skew in a static manner.
In \cite{HandlingDataSkewParallelJoinsInSharedNothingSystems}, authors addresses the issue of the redistribution skew by proposing the PRPD algorithm. However, except for redistribution skew, their approach handles the join product skew that results from the former. In PRPD algorithm, the redistribution phase of the hash-join has been modified to some degree. Especially, for the equijoin operation $R_1 \Join R_2$, the tuples of each sub-relation of $R_1$ with skewed join attribute values occurring in $R_1$ are kept locally in the database processor. On the other hand, the tuples that have skewed values happening in $R_2$ are broadcast to all the database processor. The remaining tuples of sub-relation are hash redistributed. The tuples of each sub-relation of $R_2$ are treated in the respective way. The algorithm captures efficiently the case where some values are skewed in both relations.
Using the notion of the splitting values stored in a split vector, virtual processor partitioning \cite{PracticalSkewHandlingParallelJoins} assigns multiple range partitions instead of one to each processor. Finally, authors in \cite{SkewInsensitiveParallelAlgorithmsForRelationalJoin} assign a work weight function to each join attribute value in order to generate partitions of nearly equal weight.

Finally, OJSO algorithm \cite{EfficientOuterJoinDataSkewHandlingParallelDBMS} handles data skew effect in an outer join, which is a variant of the equijoin operation.

\section{Two Motivating Examples} \label{ExamplesSection}
\label{ExamplesSection}
\noindent We will assume the simple case of a binary join operation $R_1(A,B) \Join R_2(B,C)$, in which the join predicate is of the form $R_1.B=R_2.B$. The $m$ discrete values $b_1, b_2, \ldots, b_m$ define the domain $D$ of the join attribute $B$. Let $f_i(b_j)$ denote the relative frequency of join attribute value $b_j$ in relation $R_i$. Given the relative frequencies of the join attribute values $b_1, b_2, \ldots, b_m$, the join selectivity of $R_1 \Join R_2$ is equal to \cite{OnTheRelativeCostOfSamplingForJoinSelectivityEstimation}

\begin{equation}\label{eq1}
    \mu = \sum_{b_j \in D}\prod_{i=1}^{2} f_i(b_j) = \sum_{b_j \in D} f_1(b_j)f_2(b_j)
\end{equation}

Since $\mu=\frac{|R_1 \Join R_2|}{|R_1 \times R_2|}$ and the size of the result set of the cross product $R_1 \times R_2$ is equal to the product $|R_1| |R_2|$, the cardinality of the result set associated with the join operation $R_1 \Join R_2$ is determined by the magnitude of the join selectivity.

By extending the previous analysis, the join selectivity $\mu$ can be considered as the probability of the event that two randomly picked tuples, belonging to the relations $R_1$ and $R_2$ respectively, join on the same join attribute value. Based on this observation an analytical formula concerning the size of the result set of the chain join (which is one of the most common form of the join operation) is proven. Especially we state that the join selectivity of the chain join, denoted by $R=\Join_{i=1}^{k}R_{i}(A_{i-1}, A_{i})$, is equal to the product of the selectivities $\mu_{i,i+1}$ of the constituent binary operation $R_{i}(A_{i-1}, A_{i}) \Join R_{i+1}(A_{i}, A_{i+1})$ under a certain condition of independence. In our notation, we omit to include attributes in the relations that do not participate in the join process. Formally, we have the following\\

\indent \textbf{Lemma} \emph{
Given that the values of the join attributes $A_{i}$ in a chain join of $k$ relations are independent of each other, the overall join selectivity of the chain join, denoted by $\mu$, is equal to the product of the selectivities of the constituent binary join operations, i.e., $\mu = \prod_{i=1}^{k-1}\mu_{i,i+1}$.}\\

\begin{myProof}
We define a pair of random variables $(\mathsf{X}_i, \mathsf{Y}_i)$ for every relation $R_i$, where $i=2,\ldots,k-1$. Specifically, the random variable $\mathsf{X}_i$ corresponds to the join attribute $R_i.A_i$ and it is defined as the function $\mathsf{X}_i(t) : \Omega_{i} \rightarrow \mathbb{N}_{\mathsf{X}_i}$, where $\Omega_{i}$ is the set of the tuples in the relation $R_i$. $\mathbb{N}_{\mathsf{X}_i}$ stands for the set $\{0, 1, \ldots ,|D_{A_i}|-1\}$, where $D_{A_i}$ is the domain of the join attribute $A_i$. In other words, $\mathbb{N}_{\mathrm{X}_i}$ defines an enumeration of the values of the join attribute $A_i$, in such a way that there is a one-to-one correspondence between the values of the set $D_{A_i}$ and $\mathbb{N}_{\mathsf{X}_i}$. Similarly, the random variable $\mathsf{Y}_i(t) : \Omega_{i} \rightarrow \mathbb{N}_{\mathsf{Y}_i}$ corresponds to the join attribute $A_{i+1}$, where $\mathbb{N}_{\mathsf{Y}_i}$ represents the set $\{0, 1, \ldots ,|D_{A_{i+1}}|-1\}$. \par
As for the edge relations $R_1$ and $R_k$, only the random variables $\mathsf{Y}_1$ and $\mathsf{X}_k$ are defined, since the attributes $R_1.A_0$ and $R_k.A_k$ do not participate in the join process. \par
Let $\mathcal{R}$ denote the event of the join process. Then we have that
$$p(\mathcal{R}) = p \bigl( \mathsf{Y}_1 = \mathsf{X}_2 \wedge \mathsf{Y}_2 = \mathsf{X}_3  \wedge \ldots \wedge  \mathsf{Y}_{k-1} = \mathsf{X}_k \bigr)$$

By assumption, the random variables are independent of each other. Thus, it is valid to say that
$$p(\mathcal{R}) = \prod_{i=1}^{k-1}p( \mathsf{Y}_i = \mathsf{X}_{i+1} )$$
Moreover, $p(\mathsf{Y}_i = \mathsf{X}_{i+1})$ represents the probability of the event that two randomly picked tuples from relations $R_i$ and $R_{i+1}$ agree on their values of the join attribute $A_i$. Since it holds that $p( \mathsf{Y}_i = \mathsf{X}_{i+1} ) = \mu_{i,i+1}$, the lemma follows.
\end{myProof}

\indent As a direct consequence of the previous lemma, the cardinality of the result set associated with the join operation $R=\Join_{i=1}^{k}R_{i}(A_{i-1}, A_{i})$ is given by the formula
$$|R|=\bigl( \prod_{i=1}^{k-1}\mu_{i,i+1} \bigr) \cdot \bigl( \prod_{j=1}^{k} |R_j|\bigr)$$

\subsection{Homogeneous Input Relations} \label{HomogeneousInputRelations}
\noindent Firstly, we examine the natural join of two homogeneous relations $R_1(A,B) \Join R_2(B,C)$ in the context of the join product skew effect. In the case of the homogeneous relations the distribution of the join attribute values $b_i$ is the same for both input relations $R_1$ and $R_2$. That is, there exists a distribution $f$ such that $f_1(b)=f_2(b)=f(b)$ for any $b \in D$. In this setting, the distribution $f$ is skewed when there are join attribute values $b_i,b_j \in D$ such that $f(b_i) \gg f(b_j)$.

The join attribute values with the same relative frequency $f_k$ defines the \textit{frequency class} $C_k=\{b \in D \; | \; f(b) = f_k\}$.

Thus, the domain $D$ of the join attribute $B$ is disjointly separated into classes of different frequencies. This separation can be represented with a two level tree, called \textit{frequency tree}. The nodes of the first level correspond to classes of different frequencies. The $k^{th}$ node of the first level is labeled with $C_k$. The descendant leaves of the labeled node $C_k$ correspond to the join attributes belonging to class $C_k$. Each leaf is labeled with the value of one of the join attributes of the class corresponding to the parent node.
The following picture depicts the structure of a simple frequency tree for join operation $R_1 \Join R_2$ assuming that $D = \{b_1, \ldots, b_6\}$ is separated into four frequency classes $C_1, \ldots, C_4$.


\begin{figure}
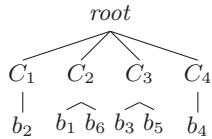

\centering
\leaf{A}
\leaf{$b_2$}
\branch{1}{$C_1$}
\leaf{$b_1$} \leaf{$b_6$}
\branch{2}{$C_2$}
\leaf{$b_3$} \leaf{$b_5$}
\branch{2}{$C_3$}
\leaf{$b_4$}
\branch{1}{$C_4$}
\branch{4}{\textit{root}}
\qobitree
\caption{The frequency tree for $R_1 \Join R_2$.}
\label{FrequencyTree}
\end{figure}
The number of produced joined tuples for a given class $C_k$ is equal to $|C_k|f_{k}^{2}|R_1||R_2|$ since $f_{k}|R_1|$ tuples of relation $R_1$ matches with $f_{k}|R_2|$ tuples of relation $R_2$ on any join attribute value $b \in C_k$. Let $N$ be the number of the database processors participating in the computation of the join operation. Since only the join product skew effect is considered, the workload associated with each node is determined by the size of the partial result set that is computed locally.
In order the workload of the join operation to be evenly apportioned on the $N$ database processors, each node should produce approximately
$\bigl(\frac{\sum_{k=1}^{K}|C_k|f_{k}^{2}}{N} \bigr)|R_1||R_2|$ number of joined tuples, where $K$ denotes the number of frequency classes. In terms of the frequency classes, this is equivalent to an appropriate assignment of either entire or subset of frequency class(es) to each database processors in order to achieve the nearly even distribution of the workload. This assignment can be represented by the selection of some internal nodes and leaves in the frequency tree. By construction, the selection of an internal node in the frequency tree amounts to the exclusive assignment of the corresponding frequency class to some database processor. Thus, this database processor will join tuples from the relations $R_1$ and $R_2$ whose join attribute value belongs to the selected class. Finally, to guarantee the integrity of the final result set, the sequence of selections must span all the leaves of the frequency tree.

\subsection{Heterogeneous Input Relations}
\noindent We extend the previous analysis in the case of heterogenous input relations. The join attribute values are distributed to the input relations $R_1(A,B)$ and $R_2(B,C)$ according to the data distributions $f_1$ and $f_2$, respectively. In general, it holds that the relative frequencies of any join attribute value $b \in D$ are different in the relations $R_1$ and $R_2$, i.e., $f_1(b) \neq f_2(b)$ for any $b \in D$. The above are depicted in table \ref{generalbinaryjoin}.

The number of joined tuples corresponding to the join attribute value $b \in D$ is rendered by the product $f_1(b) f_2(b)$. Thus, the join product skew happens when $f_1(b_i) f_2(b_i) \gg f_2(b_j) f_2(b_j)$ for some $b_i, b_j \in D$. This means that the workload of the join process for the database processor, to which the tuples with join attribute value equal to $b_i$ have been shipped at the redistribution phase, will be disproportional compared with the respective workload of another database processor. Similarly to section \ref{HomogeneousInputRelations}, the classes $C_k=\{b \in D \; | \; f_1(b) f_2(b) = f_k\}$ disjointly partition the join attribute values.

Alternatively, it is possible the definition of classes of ranges of frequencies according to the schema $C_k=\{b \in D \; | \; f_{k-1} \leq f_1(x)f_2(x) < f_k\}$ (range partitioning in the frequency level).

The ``primary-key-to-foreign-key'' join consists a special case of heterogeneity where in one of the two relation, say $R_1$, two different tuples always have different values in the attribute $B$. This attribute is called primary key and its each value $b \in D$ uniquely identifies a tuple in relation $R_1$. As to relation $R_2$, attribute $B$, called foreign key, matches the primary key of the referenced relation $R_1$. In this setting, which is very common in practice, we have that
$f_1(b_i) = \frac{1}{m}$ for any $b_i \in D$, and in general $f_2(b_i) \neq \frac{1}{m}$ with $f_2(b_i)>0$. The join product skew happens when $f_2(b_i) \gg f_2(b_j)$ for some $b_i,b_j \in D$, since $f_1(b_i)=f_1(b_j)$. Thus, the separation of the join attribute values into disjoint frequency classes can be defined with respect to the data distribution $f_2$, i.e., $C_k=\{x \in D \; | \; f_2(x) = f_k\}$.

\begin{table}
\centering
{\footnotesize
 \begin{tabular}{c||c|c}
    Join Attribute Values &  $R_1$ & $R_2$ \\ \hline \hline
  \rule[-.5cm]{0cm}{1cm} $b_1$    & $f_1(b_{1})$    & $f_2(b_{1})$ \\ \hline
  \rule[-.5cm]{0cm}{1cm} $\ldots$    & $\ldots$    & $\ldots$ \\ \hline
  \rule[-.5cm]{0cm}{1cm} $b_{m}$   &  $f_1(b_{m})$    & $f_2(b_{m})$ \\ \hline \hline
 \end{tabular}
 \vspace{0.15cm}
}
\caption{Relative frequencies of the join attribute values.}
\label{generalbinaryjoin}
\end{table}

\section{Algorithm HJPS} \label{AlgorithmSection}
In this section, we propose an algorithm, called HJPS, that alleviates the join product skew effect. The algorithm deals with the case of the binary join operation $R(A,B) \Join S(B,C)$ in which the join predicate is $R.B = S.B$.

Let $D = \{b_1, b_2, ..., b_m\}$ be the domain of values associated with the join attribute $B$. We denote by $|R_{b_{i}}|$ ($|S_{b_{i}}|$) the number of tuples of the relation $R$ (respectively $S$) with join attribute value equal to $b_i$, where $b_i \in D$. The algorithm considers that the quantities $|R_{b_{i}}|$, $|S_{b_{i}}|$ for every $b_i \in D$ are known in advance by either previously collected or sampled statistics. We also denote by $n$ the number of the database processors. In our setting, all the database processors are supposed to have identical configuration.

As it has been mentioned earlier, the number of the needed computations for the evaluation of the join operation, that identifies the total processing cost ($TPC$), is determined by the sum of products of the number of tuples in both relations that have the same join attribute values. This means that $TPC$ is expressed by the equation
$$TPC = \sum_{b_i \in D }|R_{b_{i}}|*|S_{b_{i}}|$$
\noindent In the context of the parallel execution of the join operator, the ideal workload assigned to each processor, denoted by $pwl$, is defined as the approximate number of the joined tuples that it should produce in order not to experience the join product skew effect. Obviously, it holds that that $pwl = TPC/n$.

HJPS determines whether or not a join attribute value $b_i \in D$ is skewed by the number of the processors dedicated to the production of the joined tuples corresponding to this value. To be more specific, the quotient of the division of the number of joined tuples associated with the join attribute value $b_i$ (which is equal to $|R_{b_{i}}|*|S_{b_{i}}|$) by $pwl$ gives the number of the processors needed to handle this attribute value. In the case that the result of the division, denoted by $vwl_{b_i}$, exceeds the value of two, the algorithm considers the join attribute value as skewed. The latter is inserted into a set of values, denoted by $SK$.

Let $SK=\{b_{a_{1}}, b_{a_{2}}, b_{a_{3}}, ..., b_{a_{l}}\}$ be the set of the skewed values. The algorithm iterates over the set $SK$. In particular, for the value $b_{a_{1}}$, suppose that the number of the needed processors is equal to $vwl_{b_{a_{1}}}$. The algorithm takes a decision based on the number of tuples with join attribute value $b_{a_{1}}$ in relations $R$ and $S$. If $|R_{b_{a_{1}}}|>|S_{b_{a_{1}}}|$, the tuples of the relation $R$ are redistributed to the first $vwl_{b_{a_{1}}}$ processors while all the tuples from the second relation are duplicated to all of the $vwl_{b_{a_{1}}}$ processors. In order to decide which of the $vwl_{b_{a_{1}}}$ processors is going to receive a tuple of the relation $R$ with join attribute value $b_{a_{1}}$, the algorithm applies a hash function on a set of attributes. On the contrary, if it holds that $|R_{b_{a_{1}}}|<|S_{b_{a_{1}}}|$, all the tuples from the relation $R$ with join attribute value equal to $b_{a_{1}}$ are duplicated to all of the $vwl_{b_{a_{1}}}$ processors while the tuples of the relation $S$ are distributed to all of the $vwl_{b_{a_{1}}}$ processors according to a hash function. The same procedure takes place for the rest skewed values. The remaining tuples are redistributed to the rest processors according to a hash function on the join attribute. A Pseudocode of the algorithm is given below. 

\section{Conclusion and Future Work} \label{ConclusionSection}
We address the problem of join product skew in the context of the PDBMS. In our analysis, the apriori knowledge of the distribution of the join attribute values has been taken for granted. We concentrated on the case of partitioned parallelism, according to which the join operator to be parallelized is split into many independent operators each working on a part of data. We introduced the notion of frequency classes and we examined its application in the general cases of homogeneous and heterogeneous input relations. Furthermore, an heuristic algorithmic called HJPS is proposed to handle join product skew. The proposed algorithm identifies the skew elements and assigns a specific number of processors to each of them. Given a skewed join attribute value, the number of dedicated processors is determined by the process cost for computing the join for this attribute value, and by the workload that a processor can afford.

We are looking at generalizing our analysis with frequency classes at multiway joins. In this direction we have proven the lemma of section \ref{ExamplesSection} which is about the chain join of $k$ relations. Furthermore, other types of multiway join operations, e.g., star join, cyclic join, are going to be studied in the perspective of the data skew effect and under the context of frequency classes. Finally, in a future work we will examine the case of multiway joins supposing that no statistical information about the distribution of the join attribute values is given in advance.



\fbox{
\begin{minipage}{.9\linewidth}
\begin{small}
{\em Algorithm HJPS (* Handling Join Product Skew *)} \smallskip

{\sl \textbf{Input}:}\  $t_{r_i}$ tuples of relations $R$ and $t_{r_j}$ tuples of relations $S$, $N$ number of processors. \\
{\sl \textbf{Output}:}\ correspondence of tuple to processor \\
Consider the join attribute value is the set:\\
$D = \{b_1, b_2, ..., b_m\}$\\
\hfill (* compute all frequencies for every join attribute value $in D$ *) \\
{\bf for} $j:$ $=$ $b_1$ {\bf to} $b_m$ {\bf do} \\
\tab calculate the frequencies $f_{R_j}$, $f_{S_j}$;\\
$TPC = \sum_{b_i \in D }|R_{b_{i}}|*|S_{b_{i}}|$ (*$TCP$ the total process cost*)\\
$pwl=TPC/N$ \\
(*$pwl$ the process cost of each processor*)\\
$vwl_{b_i}=|R_{b_{i}}|*|S_{b_{i}}|$;\\
(*$vwl_{b_i}$ the process cost for each join attribute value $b_i$)\\
$pn_{b_i}=vwl_{b_i}/pwl$; \\
(*$pn_{b_i}$ ideal number of processors for the join attribute value $b_i$*)\\
{\bf if} ($pn_{b_i}>=2$)
\tab consider $b_i$ a skewed value;\\
Let $SK=\{b_{a_1}, b_{a_2}, b_{a_3}, ..., b_{a_l}\}$ be the set of skewed values\\
{\bf for} $i:$ $=$ $a_1$ {\bf to} $a_l$ {\bf do} \\
\tab if $|R_{b_{a_{1}}}|>|S_{b_{a_{1}}}|$ \\
\tab\tab distribute every $t_{r_i}$ to the next $vn_{b_i}$ processors;\\
(*for distribution use a hash function to a set of attributes*)\\
\tab\tab send every $t_{s_i}$ to the next $vn_{b_i}$ processors;\\
\tab else \\
\tab \tab distribute every $t_{s_i}$ to the next $vn_{b_i}$ processors; \\
\tab\tab send every $t_{r_i}$ to the next $vn_{b_i}$ processors;\\
assign rest tuples from both relations to the rest processors; \\
(*for assignment HJPS applies a hash function to the join attribute *)\\

\end{small}
\label{fig:HJPS}
\end{minipage}
}

\end{document}